\documentclass[12pt]{article}
\usepackage[T1]{fontenc} 
\usepackage{mathtools}
\usepackage{hyperref}

\begin{document}
\date{\vspace{-5ex}}
\title{\boldmath Mass of the Stabilized Radion in the Limit of Finite Quartic Coupling}
\author{{Ali Tofighi $^{a}$\thanks{Email: a.tofighi@umz.ac.ir},\hspace{1mm} }
  and \hspace{1mm}Vahid Reza Shajiee $^{b}$
\thanks{Email: v.shajiee@mshdiau.ac.ir}\hspace{1mm} \\
{$^{a}$ \emph{Department of Physics, Faculty of Basic Science, University of Mazandaran},}\\
{\emph{P .O . Box 47416-95447, Babolsar, Iran}}\\
{$^{b}$ \emph{Young Researchers and Elite Club, Mashhad Branch, Islamic Azad University},}\\
{\emph{Mashhad, Iran}}\\  } \maketitle

\begin{abstract}

 We present an exact treatment of the modulus stabilization condition with
the general boundary conditions of the bulk scalar field in the $Randall-Sundrum$ model. We find
analytical expressions for the value of the modulus and the mass of
the radion.
 \end{abstract}

\section{Introduction}

 The hierarchy problem is one of the most attractive open problems in the modern physics. Roughly speaking, the problem is the large discrepancy between the weak and the Planck scale. The problem was addressed by several theories, like supersymmetry and higher dimensional theories, however, it has remained unsolved in the literature. Through the efforts in this direction, an impressive work is $Randall-Sundrum$ (RS) model $[1]$ which introduces a small extra dimension. The model has two branes which are called $``Planck"$ and $``TeV"$ branes, also it is assumed that a slice of $AdS_{5}$ spactime exists between the branes. A five dimensional solution to the Einstein field equations in $RS$ model is given by
\begin{equation}
ds^2=e^{-2 k r_{c} |\phi|}\eta_{\mu\nu}dx^\mu dx^\nu-r_{c}^2 d\phi^2,
\end{equation}
where $-\pi \leq \phi \leq \pi$ is the extra dimension coordinate, the coefficient $r_{c}$ is
the compactification radius, and the parameter k is related to the $5$-D Planck mass, $M$.

It would be interesting to check out the stability of the extra dimension, $\phi$, in the RS model. Such an investigation was addressed by $Goldberger$ and $Wise$ (GW) $[2]$. The $GW$ model contains a massive scalar field with usual kinetic term in the bulk and quartic interactions localized on the branes. The original work of Ref. $[2]$ has several deficiencies, namely:\\
1) The bulk mass term breaks the conformal invariance of the
theory.\\
2) In the limit of infinite quartic coupling considered in Ref.
$[2]$, it is not possible to unravel the complete structure of the
critical points of the theory, hence they miss the source of
instability as indicated
by the existence of a closely spaced maximum.\\
3) The boundary conditions of the model comprised of a pair of
coupled cubic algebraic equations. In Ref. $[2]$ these boundary
conditions are not solved, instead
they merely choose a specific configuration.\\
4) In addition they only consider the
 leading order term in their calculations.
Hence their treatment of the subject matter is an approximate one.\\
Therefore, many studies were done on this subject $[3-12]$.
In $[3-5]$, the authors have considered the stabilization
of the modulus containing a scalar field, which interacts with the
spacetime curvature R, in the bulk. These theories have conformal invariance at
certain value of the coupling constant of the curvature and the
scalar field,
hence they remedy the objection $1$, raised above.

 An exact analysis of the $GW$ mechanism has been discussed in Ref.
$[6]$ and the objections $2-4$ have been addressed. However their
treatment of the stabilized modulus is a numerical one. They also do
not address the issue of the mass of radion at finite values of the
quartic coupling constant. The issue of Goldberger-Wise mechanism with the general boundary
conditions also has been discussed in Ref. $[13]$. Recently, the issue of stability of the Randall-Sundrum model has been discussed in the framework of $AdS/CFT$ correspondence $[15][16]$.

 The motivation for the present study is to discuss the the relevance of going
beyond the infinite quartic coupling limit, in physical terms
and to explore it's phenomenological implications. The plan of this paper as follows:
 In section two, we describe the model and calculate the effective potential, its extremization condition and the value of the stabilized modulus. In section three,
we study the modulus for the case where the quartic coupling is
finite but very large. we obtain the mass of the radion in this
limit as well. Finally in section four we present our conclusions.

\section {Review of the GW mechanism}
In this section, mainly based on [6], we present a brief review of the GW mechanism in the limit of finite quartic coupling. The action of the model is  of the form:
\begin{equation}
S=S_{gravity}+S_{vis}+ S_{hid}+ S_{\Phi},
\end{equation}
where,
\begin{equation}
S_{gravity}=\int d^4x\int^\pi_{-\pi}d\phi\sqrt{G}[2M^3R-\Lambda],
\end{equation}
 \begin{equation}
 S_{vis}=\int d^4x\sqrt{-g_s}[L_s-V_s],\qquad
 S_{hid}=\int d^4x\sqrt{-g_p}[L_p-V_p],
 \end{equation}
\begin{eqnarray}
S_{\Phi}&=&\frac{1}{2}\int
dx^4\int^\pi_{-\pi}d\phi\sqrt{G}(G^{MN}\partial_M\Phi\partial_N\Phi
-m^2 \Phi^2) -\int
 dx^4\sqrt{-g_s}\lambda_s(\Phi^2-v^2_s)^2\nonumber\\
 &&-\int
dx^4\sqrt{-g_p}\lambda_p(\Phi^2-v^2_p)^2,
\end{eqnarray}
where  $\Lambda$ is the five dimensional cosmological constant,
$V_s$,$V_p$ are the visible and hidden brane tensions. And
$G=det[G_{MN}]$.\\
The $\phi$-dependent vacuum expectation value $\Phi(\phi)$ is
obtained
 from the equation of motion
\begin{equation}
\partial_\phi(e^{-4\sigma}\partial_\phi \Phi)=m^2
r^2e^{-4\sigma}\Phi+ 4e^{-4\sigma}\lambda_s r
\Phi(\Phi^2-v^2_s)\delta(\phi-\pi)+4e^{-4\sigma}\lambda_p r
\Phi(\Phi^2-v^2_p)\delta(\phi),
\end{equation}
where $\sigma = k r_{c} | \phi |$. Away from the boundaries ($\phi=0,\pi$) the solution is
\begin{equation}
\Phi(\phi)=Ae^{(\nu+2)\sigma}+Be^{(-\nu+2)\sigma},
\end{equation}
where $\nu=\sqrt{4+\frac{m^2}{k^2}}\approx 2+\epsilon$.\\ If we
insert this solution in $eq. (5)$ and integrate over $\phi$ we
obtain the effective $4-$ dimensional potential for the modulus $r$,
namely $V_{\Phi}(r)$ which is given by
\begin{eqnarray}
 V_{\Phi}(r)&=&k(\nu+2)A^2(e^{2\nu k r\pi}-1)+
        k(\nu-2)B^2(1-e^{-2\nu k r\pi})+
\lambda_se^{-4 k r\pi}(\Phi^2(\pi)-v^2_s)^2\nonumber\\
&&+\lambda_p(\Phi^2(0)-v^2_p)^2.
\end{eqnarray}
 The coefficients A and B are determined by
imposing appropriate boundary conditions on the 3-branes. Putting $eq. (7)$ into $eq. (6)$ and matching delta functions, the conditions are obtained as
\begin{equation}
k[(2+\nu)A + (2-\nu)B] -2\lambda_{p}\Phi(0)[\Phi(0)^2 - v_{p}^2]=0,
\end{equation}
\begin{equation}
k e^{2kr\pi}[(2+\nu)e^{\nu kr\pi}A + (2-\nu)e^{-\nu kr \pi}B] +
2\lambda_{s}\Phi(\pi)[\Phi(\pi)^2 - v_{s}^2] = 0.
\end{equation}
For arbitrary value of $\lambda$ the boundary values of the
scalar field at the two orbifold fixed points are $\Phi(\phi = 0) =
Q_p(r)$ and $\Phi(\phi = \pi) = Q_s(r)$. Now A and B can be written, from $eq. (7)$, in terms of boundary values of the scalar field as follows
\begin{equation}
A = \frac {Q_s(r) e^{-2 \sigma} - Q_p(r) e^{- \nu \sigma}}{ 2 \sinh(\nu \sigma)},
\end{equation}
\begin{equation}
B = \frac {Q_p(r) e^{\nu \sigma} - Q_s(r) e^{- 2 \sigma}}{ 2 \sinh(\nu \sigma)}.
\end{equation}
Putting above expressions for A and B into $eq. (9)$ and $eq. (10)$, we get
\begin{equation}
\frac{\nu}{2sinh(\nu \sigma)}[e^{-2\sigma}-(
\frac{\nu+2}{2\nu}e^{-\nu\sigma}+\frac{\nu-2}{2\nu}e^{\nu
\sigma})\frac{Q_p}{Q_s}]=\frac{2\lambda_p}{k}\frac{Q_p}{Q_s}(Q_p^2-v_p^2),
\end{equation}
and
\begin{equation}
\frac{\nu}{2sinh(\nu \sigma)}[\frac{Q_p}{Q_s}-(
\frac{\nu+2}{2\nu}e^{(\nu-2)\sigma}+\frac{\nu-2}{2v}e^{-(\nu+2)\sigma})]
=\frac{2\lambda_s}{k}(Q_s^2-v_s^2)e^{-2\sigma}.
\end{equation}
Using $eq. (11)$, $eq. (12)$ and $eq. (14)$ into the extremization condition for the effective potential ($\frac{dV_{\Phi}(r)}{dr}=0$), the modulus can be obtained
\begin{equation}
kr_{\pm}=\frac{1}{\pi(\nu-2)}\ln[\frac{1}{(\frac{2+\nu}{2\nu}+\frac{\nu-2}{2\nu}e^{-2\nu\sigma})}
(\frac{Q_p(r)}{Q_s(r)}) (\frac{1}{1\pm C\sqrt{\frac{\lambda_s
Q^2_s(r)}{1+\lambda_sQ^2_s(r)}}})],
\end{equation}
where
\begin{equation}
C=\sqrt{1-\frac{4[(\nu+2)e^{2(\nu-2)\sigma}-e^{-4\sigma}(4-\nu^2)+(2-\nu)e^{-2(\nu+2)\sigma}]}
{[(2+\nu)e^{(\nu-2)\sigma}+(\nu-2)e^{-(\nu+2)\sigma}]^2}}.
\end{equation}
For $kr_{+}$, $\frac{d^{2}V}{dr^{2}}>0$, and for $kr_{-}$, $\frac{d^{2}V}{dr^{2}}<0$. So, $kr_{+}$ and $kr_{-}$ are minimum and maximum of the potential, respectively. Clearly, $kr_{+}$ and $kr_{-}$ are respectively correspond to stability and instability of the modulus field or the radion field.

\section{The modulus and the mass of radion}
In this section we consider some observable quantities such as the
modulus and the mass of radion.

\subsection{ The modulus}
Now we study the stable and unstable values of the modulus for the original $GW$ mechanism at finite coupling. The analytic expression for the modulus was obtained and analyzed in Ref. $[6]$, but we want to investigate the modulus by taking the parameter $\epsilon$ as a variable. In the large kr limit, the values of the stable and unstable modulus when
$\lambda_p\rightarrow\infty$,$\lambda_s\rightarrow\infty$ are
\begin{equation}
kr_{\pm}=\frac{1}{\pi(\nu-2)}\ln[ \frac{2\nu
v_p}{(\nu+2\pm\sqrt{\nu^2-4})v_s}].
\end{equation}
In order to consider the value of the stabilized modulus for the
large but finite value of the quartic coupling constant, a $\frac{1}{\lambda}$ expansion of boundary scalar field should be considered
\begin{equation}
Q_p(r)=v_p+\frac{k}{\lambda_p v_p}\frac{\nu e^{-2\sigma}}{4\sinh(\nu
\sigma)}[\frac{v_s}{v_p}-(\frac{2+\nu}{2\nu}e^{(2-\nu)\sigma} +
\frac{\nu-2}{2\nu}e^{(\nu+2)\sigma})],
\end{equation}
\begin{equation}
Q_s(r)=v_s+\frac{k}{\lambda_s v_s}\frac{\nu e^{2\sigma}}{4\sinh(\nu
\sigma)}[\frac{v_p}{v_s}-(\frac{2+\nu}{2\nu}e^{(\nu-2)\sigma} +
\frac{\nu-2}{2\nu}e^{-(\nu+2)\sigma})].
\end{equation}
The values of the modulus at finite value of the quartic coupling are $[6]$
\begin{equation}
kr_{\pm}=\frac{1}{\pi(\nu-2)}ln[\frac{2\nu}{2+\nu}\frac{n}{1\pm
\sqrt{\frac{v-2}{v+2}}(1-\frac{q}{2})}
(1-\frac{t(\nu-2)}{4}+\frac{q(\nu+2)}{4}-\frac{q\nu n}{2}
e^{(2-\nu)k \pi r_{\pm}})],
\end{equation}
where
\begin{equation}
 n=\frac{v_p}{v_s},\qquad t=\frac{k}{\lambda_pv^2_p},\qquad
q=\frac{k}{\lambda_sv^2_s}.
\end{equation}
The values of the stable and unstable modulus $kr_{\pm}$ are obtained by solving
$eq.(20)$ which are
\begin{equation}
kr_{\pm}=\frac{1}{\pi(\nu-2)}\ln[\frac{\alpha+\sqrt{\alpha^2-4\beta}}{2}],
\end{equation}
where
\begin{eqnarray}
\alpha&=&\gamma_{\pm}(1-\frac{t(\nu-2)}{4}+\frac{q(\nu+2)}{4}),\qquad
\beta=
\frac{\gamma_{\pm}\nu q n}{2},\qquad with \qquad \nonumber\\
\gamma_{\pm}&=&\frac{2\nu
n}{\nu+2}\frac{1}{1\pm\sqrt{\frac{v-2}{v+2}}(1-\frac{q}{2})}.
\end{eqnarray}
Using above equation, we can analyze the modulus $kr_{\pm}$ by taking the parameter $\epsilon$ as a variable. Figure $1$ shows the variation of the stable modulus $kr_{+}$ versus the
parameter $\epsilon$. In this figure the value of $v_s=1$ and
$v_p=1.2$. The solid curve corresponds to case of infinite coupling.
The dashed curve corresponds to the case where $t=q=0.1$ and the
dotted curve corresponds to the case where $t=q=0.2$. As seen above, $kr_{-}$ is another value of the modulus which, because of $\frac{d^{2}V}{dr^{2}}<0$, corresponds to the maximum of the potential. Due to the maximality, $kr_{-}$ leads to instability, i.e. for any small perturbations the system will roll down to the minimum. Figure $2$ shows the variation of the unstable modulus $kr_{-}$ versus the parameter $\epsilon$. The value of $v_s=1$ and
$v_p=1.2$. The solid curve corresponds to case of infinite coupling.
The dashed curve corresponds to the case where $t=q=0.1$ and the
dotted curve corresponds to the case where $t=q=0.2$. From figures $1$ and $2$, we observe that by increasing the value of the parameters $t$ and $q$, we obtain more deviation from usual Randall-Sundrum case. As we know the value of $kr_{+}\sim12$ may solve hierarchy problem. From this point of view, the figure $1$ shows that the parameter $\epsilon$ in the finite coupling case can be smaller in comparison with the infinite coupling case.

It is appropriate to study the difference of the modulus at finite
and infinite values of the quartic coupling constant, the result is
\begin{equation}
4\pi
[(kr_+)_\infty-(kr_+)_f]=\frac{q}{\sqrt{\epsilon}}+\frac{1}{2}(q+2t)+
\frac{\sqrt{\epsilon}q}{8}+\ldots,
\end{equation}
where in the series expansion we have kept the linear terms of the
parameters $t$ and $q$.\\

\begin{center}
  \includegraphics[scale=0.60]{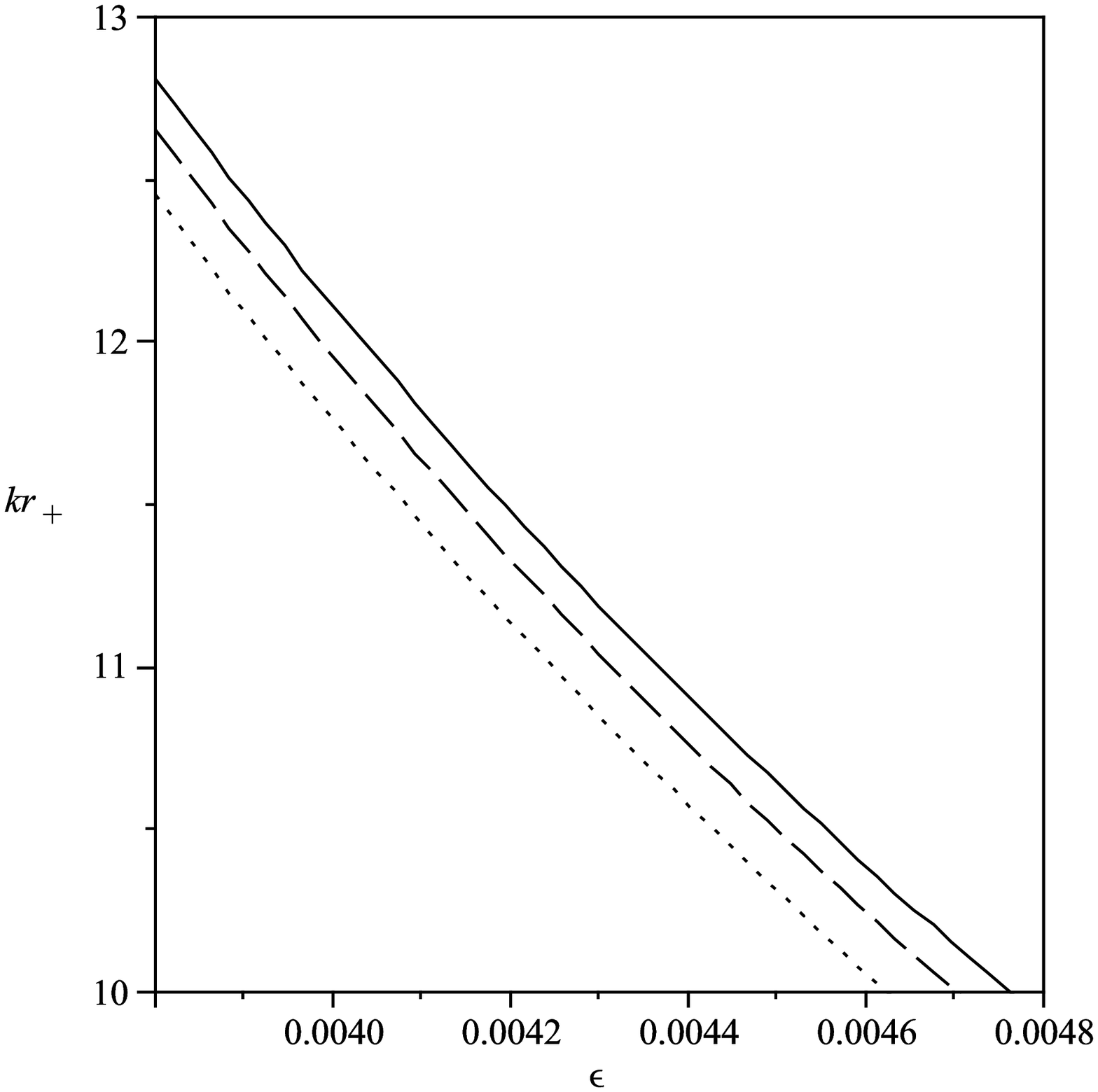}\\
  {\small {FIGURE 1: The variation of $kr_+$ defined by $eqs.(17,22)$ versus the
parameter $\epsilon$. The value of the parameters are
$v_s=1$, $v_p=1.2$. The solid curve corresponds to limit of
infinite coupling ($t=q=0.0$). For the dashed curve the value of
these parameters are $t=q=0.1$, and for the dotted curve are
$t=q=0.2$.}}\\
\end{center}

\begin{center}
  \includegraphics[scale=0.60]{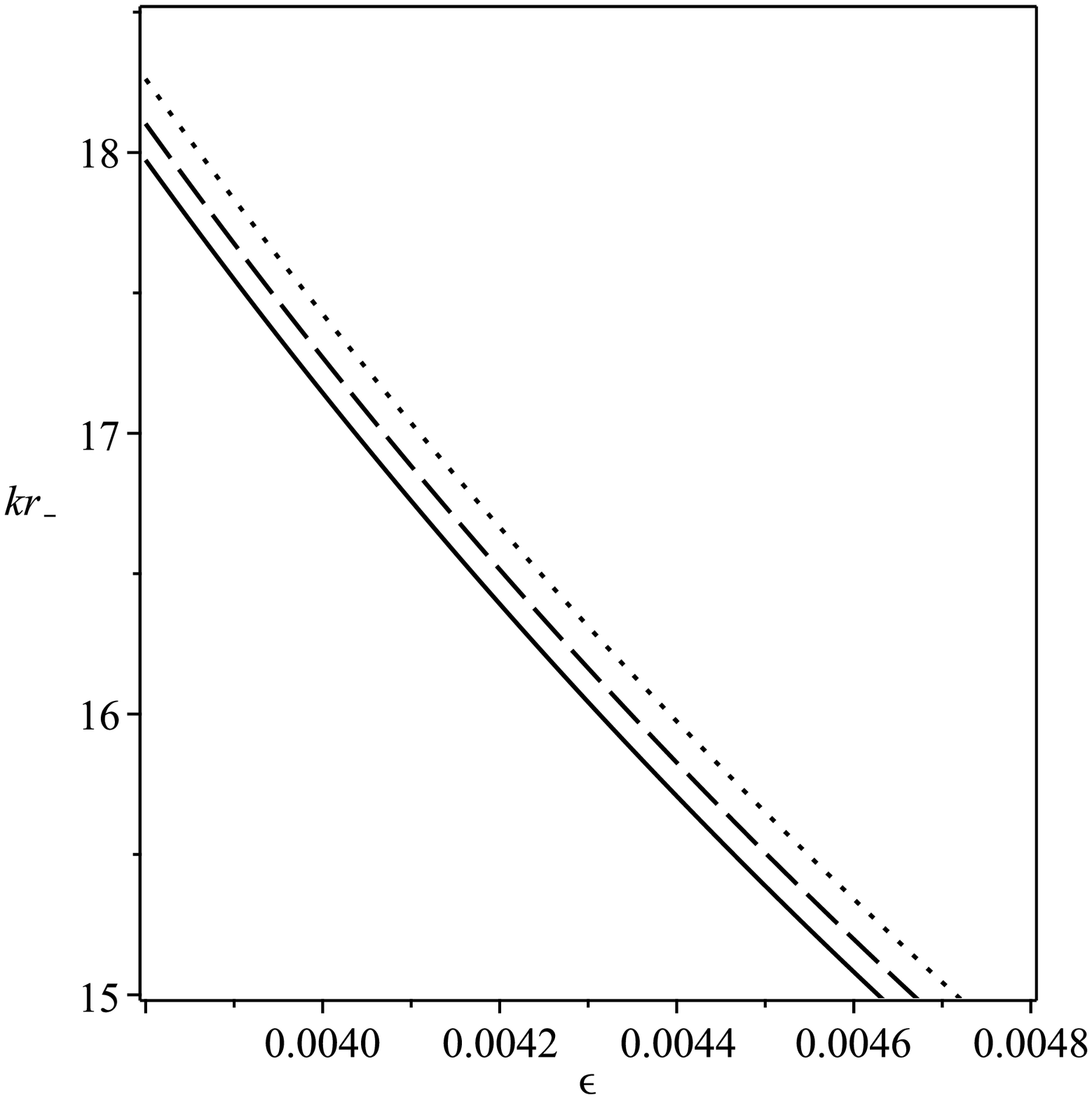}\\
  {\small {FIGURE 2: The variation of $kr_-$ defined by $eqs.(17,22)$ versus the
parameter $\epsilon$. The value of the parameters are
$v_s=1$, $v_p=1.2$. The solid curve corresponds to limit of
infinite coupling ($t=q=0.0$). For the dashed curve the value of
these parameters are $t=q=0.1$, and for the dotted curve are
$t=q=0.2$.}}\\
\end{center}

 \subsection {Mass of the radion} The
phenomenology of radion has been addressed in Refs. [8,11,14]. The
canonically normalized radion field is $\Phi=f \hat{\Phi}$, where $f
=\sqrt{ \frac{6M^3}{k}}$ is another scale of order of Planck mass
and
\begin{equation}
\hat{\Phi}=e^{-k\pi r}.
\end{equation}
The radion mass is defined the second derivative of $V_\Phi$ with
respect to the canonically normalized radion field evaluated at its
minimum. Hence
\begin{equation}
m^2_\Phi=\frac{d^2V_\Phi}{d\Phi^2}|_{\Phi=\Phi_+}=[(\frac{dr}{d\Phi})^2
(\frac{dV^2_{\Phi}(r)}{dr^2})]|_{r=r_+}
=\frac{1}{(fk\pi)^2}[e^{2kr\pi}(\frac{dV^2_{\Phi}(r)}{dr^2})]|_{r=r_+}
\end{equation}
In order to calculate the mass of radion, we calculate the second
derivative of the effective potential and we get
\begin{eqnarray}
\frac{dV^2_{\Phi}(r)}{dr^2}&=&-\frac{4k\pi\nu
e^{-2\sigma}}{sinh(\nu\sigma)}[\lambda_p(Q_p^2-v^2_p)Q_pQ'_s+
\lambda_s(Q_s^2-v^2_s)Q_sQ'_p
\nonumber\\
&+&2\pi\lambda_p\lambda_s(Q^2_s-v^2_s)(Q^2_p-v^2_p)Q_pQ_s].
\end{eqnarray}
In the limit of
$\lambda_p\rightarrow\infty$,$\lambda_s\rightarrow\infty$, $Q_s=v_s$
and $Q_p=v_p$,
 we find
\begin{equation}
(m^2_\Phi)_\infty=\frac{8k^2 v_p v_s \sqrt{\epsilon^3}e^{-\nu \pi
(kr_+)_\infty}}{3M^3}.
\end{equation}
In Ref. $[14]$ the scale factor $f=\sqrt{\frac{24M^3}{k}}$, moreover
in their work $e^{\pi(\nu-2) k r}=n$. Now by using these values for
the scale factor $f$ and the modulus $kr$,
 from $eqs.(22,23)$ we obtain
\begin{equation}
(m^2_\Phi)_\infty=\frac{k^2  v^2_s \epsilon^2e^{-2 k\pi r}}{3M^3},
\end{equation}
which is identical to the result of $[14]$. This in turn validates
our results for the second derivative of the effective potential
against a similar calculations presented in Ref. $[6]$.

Similarly we can calculate the mass of radion in the case where the
quartic coupling constants are finite, the result is
\begin{equation}
(m^2_\Phi)_f=\frac{8k^2 Q_p(r) Q_s(r) \sqrt{\epsilon^3}e^{-\nu \pi
(kr_+)_f}}{3M^3}.
\end{equation}
Since $k \sim M_{pl} \sim M$ and $Q_p(r) \sim M_{pl}^{3/2} \sim Q_s(r)$, the radion mass $m_{\Phi}$ at finite coupling is $\mathcal{O}(TeV)$ when $kr_{+}\sim12$. This is in agreement with previous works in the context of radion phenomenology. As seen above, the radion mass has dependency on the parameter $\epsilon\equiv m^{2}/4k^{2}$ as $\epsilon^{3/2}$. If one takes into account backreaction of
the stabilizing field $\Phi$ on the background geometry, the radion mass turns out to
have $\epsilon^{2}$ dependance $[17]$. This discrepancy
of $\epsilon$-dependence between these two approaches may come from the fact that during all calculations we have assumed that $\nu\approx2+\epsilon$ ($\epsilon\ll1$) and we have neglected higher order terms of $\epsilon$. It would be of interest to take $\nu\approx2+\epsilon-1/4\epsilon^{2}$ and study similarities and differences between the effective potential and the exact gravity-scalar approaches.

It is appropriate to study the logarithmic ratio of these masses
which we denote by $\chi$ defined by
\begin{equation}
\chi=ln[\frac{(m^2_\Phi)_f}{(m^2_\Phi)_\infty}]=\nu\pi
[(kr_+)_\infty-(kr_+)_f]=\frac{1}{2}[\frac{q}{\sqrt{\epsilon}}+\frac{1}{2}(q+2t)+
\frac{5\sqrt{\epsilon}q}{8}+\ldots].
\end{equation}

 Figure $3$ shows the variation of the  $\chi$ versus the
parameter $\epsilon$. In this figure the value of $v_s=1$, and
$v_p=1.2$.
 The dashed
curve corresponds to the case where $t=q=0.1$ and the
dotted curve corresponds to the case where $t=q=0.2$. As seen in figure $3$, the mass of the radion $(m_\Phi)_f$ at finite value of
$\lambda$ could be much larger than the original value
reported in Ref. $[14]$ (the infinite coupling case). From phenomenological point of view, this is a reasonable result because it shows that the mass of the radion is governed by the strength of radion coupling. Moreover, since the radion mass increases as the finite quartic coupling becomes smaller, it could be considered in the context of high-mass radion or Higgs-radion mixed scenarios $[18-24]$.

\begin{center}
  \includegraphics[scale=0.60]{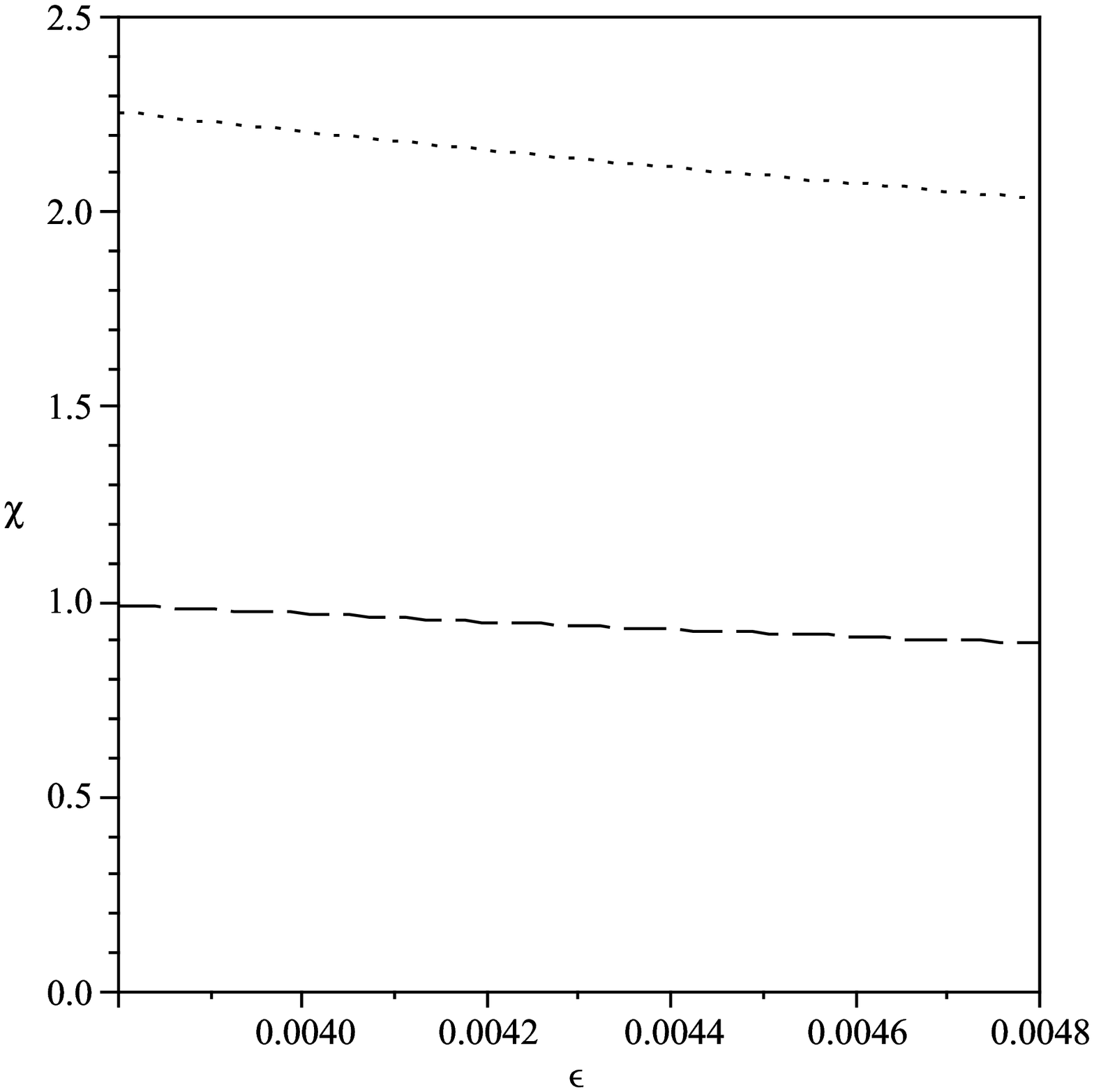}\\
  {\small {FIGURE 3: The variation of $\chi$ defined by $eq.(31)$ versus the parameter
$\epsilon$. The value of the parameters are $v_s=1$,
$v_p=1.2$. For the dashed curve the value of the parameters
are $t=q=0.1$, and for the dotted curve are $t=q=0.2$. }}\\
\end{center}

\clearpage

\section{Conclusions}

We have made a critical assessment of the $GW$ mechanism for the
stabilization of modulus. We have managed to extend the work initiated in Ref. $[6]$ to the phenomenology of radion. The limit
studied by Goldberger and Wise, because the boundary conditions
become very simple in this case corresponds to
 the limit of infinite quartic coupling of the scalar potential term on the
boundary branes. We have succeeded to refine this aspect of the mechanism.
We have found correction for the value of the modulus and the mass of the
radion. Instead of brane potential with quartic coupling, it is also
possible to consider brane potential of the quadratic form. We plan
to report on these issues in the future. As final note, it is possible to stabilize the modulus by a massless scalar non-minimally coupled to gravity. In such work, it would be of interest to study in the limit of finite coupling.

\section{Competing Interests}
The authors declare that there is no conflict of interest regarding the publication of this paper.

\section{Acknowledgment}
The authors would like to thank the anonymous referee for his/her valuable comments and suggestions to improve the quality of the paper.

\clearpage


\end{document}